\documentclass[11pt,fleqn,twoside]{article}
\usepackage{amsfonts,amssymb,latexsym}
\makeatletter
\newcommand{\prava}{\footnotesize\it
\begin{flushright}
\begin{minipage}{18cm}
Copyright \copyright 1998 by W.W. Zachary and V.M. Shtelen
\end{minipage}
\end{flushright}}

\newcommand{\name}[1]{\begin{flushleft}
                       \LARGE \bf #1
                       \end{flushleft}\vspace{-3mm}}

\newcommand{\Author}[1]{\begin{flushleft}
                       \it #1 \end{flushleft}}

\newcommand{\Adress}[1]{\begin{flushleft}
                       \it #1 \end{flushleft}}

\newcommand{\Date}[1]{\begin{flushleft}
                      \small  \it #1 \end{flushleft}}

\newcommand{\ehkol}{Author \ name}
\newcommand{\ohkol}{Article \ name}
\renewcommand{\@evenhead}{
\hspace*{-3pt}\raisebox{-15pt}[\headheight][0pt]{\vbox{\hbox to \textwidth
{\thepage \hfil \ehkol}\vskip4pt \hrule}}}
\renewcommand{\@oddhead}{
\hspace*{-3pt}\raisebox{-15pt}[\headheight][0pt]{\vbox{\hbox to \textwidth
{\ohkol \hfil \thepage}\vskip4pt\hrule}}}
\renewcommand{\@evenfoot}{}
\renewcommand{\@oddfoot}{}

\newcommand{\be}{\begin{equation}}
\newcommand{\ee}{\end{equation}}
\newcommand{\ba}{\hspace*{-5pt}\begin{array}}
\newcommand{\ea}{\end{array}}
\newcommand{\p}{\partial}
\newcommand{\ds}{\displaystyle}
\makeatother

\begin{document}
\setcounter{page}{417}
\thispagestyle{empty}

\renewcommand{\ehkol}{W.W. Zachary and V.M. Shtelen}
\renewcommand{\ohkol}{On Asymptotic Nonlocal Symmetry of Nonlinear
Schr\"odinger Equations}

\begin{flushleft}
\footnotesize \sf
Journal of Nonlinear Mathematical Physics \qquad 1998, V.5, N~4,
\pageref{zachary-fp}--\pageref{zachary-lp}.
\hfill {\sc Article}
\end{flushleft}

\vspace{-5mm}

\renewcommand{\footnoterule}{}
{\renewcommand{\thefootnote}{} \footnote{\prava}} 

\name{On Asymptotic Nonlocal Symmetry of \\
Nonlinear Schr\"odinger Equations} \label{zachary-fp}

\Author{W.W. ZACHARY~$^\dag$ and V.M. SHTELEN~$^\ddag$}

\Adress{$\dag$~Department of Electrical Engineering, Howard
University,\\ 
~~Washington, DC 20059   USA\\
~~E-mail: chris50@radix.net\\[1mm]
$\ddag$~Department of Mathematics, Hill Center, Rutgers University,\\
~~Piscataway, NJ 08854     USA\\
~~E-mail: shtelen@lagrange.rutgers.edu}

\Date{Received May 25, 1998; Accepted July 15, 1998}

\begin{abstract}
\noindent
A concept of asymptotic symmetry is introduced which is based on a
def\/inition of symmetry as a reducibility property relative to a
corresponding invariant ansatz. It is shown that the nonlocal Lorentz
invariance of the free-particle Schr\"odinger equation, discovered by
Fushchych and Segeda in 1977, 
can be extended to Galilei-invariant equations for free particles
with arbitrary spin and, with our def\/inition of asymptotic symmetry,
to many nonlinear Schr\"odinger equations. An important class of
solutions of the free Schr\"odinger equation with improved smoothing
properties is obtained.

\end{abstract}

\section{Introduction}

It is well-known that the maximal Lie invariance algebra of the free
linear Schr\"odinger equation in three spatial dimensions
\be
\left( i\hbar \p_t+\frac{\hbar^2}{2m} \Delta\right) \psi=0,
\ee
is the Schr\"odinger algebra $sch(1,3)$, a Lie algebra which contains
the Galilei and dilation algebras as well as some special conformal
transformations [1--3]. It was quite surprising therefore, when
Fushchych and Segeda [4] showed that equation~(1) is 
also invariant under an algebra of nonlocal pseudodif\/ferential
operators which is isomorphic to the Lie algebra~$so(1,3)$ of the
three-dimensional homogeneous Lorentz group.

The purpose of the present work is to establish similar results
concerning nonlocal symmetry for some other equations of interest in
mathematical physics. In particular, it will be shown that linear
Schr\"odinger equations with linear and quadratic potentials with
arbitrary time-dependent coef\/f\/icients and Hurley's Galilei-invariant
wave equations~[5], which describe free nonrelativistic quantum
particles with arbitrary spin, are invariant under related algebras
of nonlocal pseudodif\/ferential operators. The results on nonlocal
symmetries for linear Schr\"odinger equations with the potentials
described above follows from the fact that these equations can be
transformed to the free Schr\"odinger equation (1) and, therefore,
inherit the nonlocal symmetry of the latter equation. The generators
of the respective nonlocal Lie algebras for these Schr\"odinger
equations necessarily have dif\/ferent representations than 
the generators of the algebras corresponding to equation~(1), but the
algebras associated, respectively, with these dif\/ferent equations are
isomorphic. 

Analogous results to those for the linear Schr\"odinger equations
mentioned above can be obtained for some nonlinear Schr\"odinger
equations (NSEs). Specif\/ically, a subclass of the family of NSEs
proposed and discussed by Doebner and Goldin over the past several
years~[6, 7] in connection with representations of dif\/feomorphism 
groups and the corresponding Lie algebras of vector f\/ields, as well
as some related equations studied by Auberson and Sabatier~[8] have
been shown to be linearizable; i.e., they can be mapped by means of
an appropriate change of variables to linear Schr\"odinger equations.
In particular, the potential in the latter equation can be chosen to
be identically zero so that the solutions of these nonlinear
equations can be related to the nonlocal invariant solutions of
equation~(1). These results make the situation for these linearizable
NSEs analogous to the situation for linear Schr\"odinger equations
with time-dependent linear and quadratic potentials, where the
isomorphism of the respective Lie algebras was f\/irst proved by
Niederer~[9] for the linear Schr\"odinger equation with the usual
(time-independent) harmonic oscillator potential. In the cases
examined by Niederer, however, the Lie algebras were those
corresponding to the Schr\"odinger and oscillator groups whereas, in
the present paper, we consider isomorphisms for nonlocal Lie
algebras. 

Since the symmetries that we discuss are nonlocal, Lie's approach is
not adequate to deal with them. A generalization of Lie's method,
suitable for linear partial dif\/ferential equations (PDEs), was
suggested in [10] and is based upon the following commutator form of
invariance condition.

\medskip

\noindent
{\bf Def\/inition 1.1.} {\it An operator $Q$ is a symmetry of a linear
system of PDEs 
\be
Lu = 0                                       
\ee
if and only if $[L, Q]u = 0$ for each solution $u$ of (2).
}

\medskip

Note that there is no restriction on the order of the operator $Q$
(unlike the situation with Lie's methods [11]). In fact, $Q$ need not
be a dif\/ferential operator, but may also be a pseudodif\/ferential or
integral operator. This non-Lie approach proved to be very ef\/fective,
and wide classes of new symmetries were discovered for many equations
of mathematical physics [12]. 

In the present paper we suggest an alternative symmetry criterion of
invariance for nonlocal operators which is also suitable for
nonlinear equations. 

\medskip

\noindent
{\bf Def\/inition 1.2.} {\it We will say that an operator $Q$ is a symmetry of
a system of PDEs if and only if a corresponding ansatz, obtained as a
solution of the equation
\be
Qu=0,                                           
\ee
reduces the given system of PDEs.}

\medskip

\noindent
{\bf Remark.} When $Q$ is a (linear) pseudodif\/ferential operator, one
can construct an invariant ansatz (solution of (3)) by means of a
Fourier transform, as shown in [13] (see also [3], Sec. 5.11).

\medskip

We prove that certain classes of nonlinear Schr\"odinger equations
(NSEs) have a similar invariance when the time is larger than the
squared modulus of the spatial variables in a certain well-def\/ined
sense. This ``asymptotic symmetry'' is based upon Def\/inition~1.2 when
there does not exist an exact reduction of the given system of
nonlinear PDEs by an invariant ansatz, but there does exist such a
reduction in a well-def\/ined asymptotic sense. (The precise def\/inition
will be given in Section~4.) This asymptotic symmetry situation is
somewhat reminiscent of the situation encountered in scattering
theory in which a solution of a nonfree equation approaches a
solution of the corresponding free equation as $t\to \pm\infty$~(cf.
[14]). This scenario has been established for some NSEs with
power-type nonlinearities~[15]. In the present case, certain NSEs
have a nonlocal asymptotic symmetry which is an exact symmetry of the
free Schr\"odinger equation. 

As we shall discuss in more detail later, a number of authors have
given Lie symmetry analyses for the NSEs that we consider [3, 11].
However, since the symmetries of interest in the present paper are
nonlocal. Lie's algorithm cannot be applied to them and a dif\/ferent
approach must be used to study them.

The paper is organized in the following manner. In Section~2 we
discuss the results for the free Schr\"odinger equation in further
detail, extending the three-dimensional results of~[4] to any spatial
dimension $n \geq 2$ and extending the nonlocal Lorentz-invariant
solutions of the free Schr\"odinger equation to a larger class 
of symmetries. Nonlocal symmetries of the type discussed in the
present paper do not exist when $n = 1$. They are phenomena peculiar
to higher spatial dimensions. Similar results are obtained for the
three-dimensional Galilei-invariant free-particle 
wave equations of Hurley [5] for nonrelativistic free quantum
particles with arbitrary spin. We have found a new representation of
the Lorentz algebra for particles with arbitrary spin whose time
evolution is described by these equations. In particular, the spin
contributions to the angular momentum operators do not satisfy the
familiar angular momentum commutation relations, even though the
total angular momentum operators satisfy the correct commutation
relations with themselves and with the nonlocal generators.

In Section~3 we show that, since linear Schr\"odinger equations
with linear and quadratic (harmonic oscillator) potentials with
arbitrary time-dependent coef\/f\/icients can be transformed to the free
Schr\"odinger equation (1), they inherit the nonlocal symmetry of the
free equation. The corresponding nonlocal Lie algebras are
isomorphic. Similar results are then discussed for the linearizable
Doebner-Goldin and Auberson-Sabatier NSEs.

In Section~4 we prove that several classes of NSEs possess nonlocal
symmetries in an asymptotic sense when the time variable is
suf\/f\/iciently large. The collection of NSEs considered include several
classes of standard type with power-type nonlinearities as well as
the families of NSEs proposed and discussed by Doebner and Goldin
[6, 7], Bialynicki-Birula and Mycielski [16], and Kostin [17]. 
Finally, Section~5 consists of some concluding remarks.

\newpage

\section{Free-particle equations}

We f\/irst discuss the exact nonlocal symmetry of the free-particle
Schr\"odinger equation, and then consider analogous exact nonlocal
symmetries for the Galilei-invariant wave equations of Hurley.

\subsection*{Free-particle Schr\"odinger equations}

We def\/ine the standard operators $\ds p_0\equiv i\hbar \p_t$,
$p_j \equiv -i \hbar \p_{x_j}$ $(j = 1, 2,\ldots, n)$, and set $\ds
L_S\equiv p_0-\frac{p_j p_j}{2m}$ where, in the latter def\/inition,
the convention of summing over repeated indices is used. Then the
free-particle Schr\"odinger equation (1) takes the 
form $L_S\psi = 0$. Consider the operators $(j, k = 1, 2, \ldots, n
\geq 2)$: 
\setcounter{equation}{3}
\renewcommand{\theequation}{\arabic{equation}{\rm a}}
\be
J_{jk} =x_j p_k -x_k p_j,
\ee
and
\setcounter{equation}{3}
\renewcommand{\theequation}{\arabic{equation}{\rm b}}
\be
J_{0j}=\frac{1}{2m} (f(p)G_j +G_j f(p)),
\ee
where     
\setcounter{equation}{3}
\renewcommand{\theequation}{\arabic{equation}{\rm c}}
\be
G_j =t p_j -m x_j,
\ee
$p \equiv \left(\sum\limits_{j=1}^n p_j p_j\right)^{1/2}$, and $f$ denotes a
smooth function.

One easily verif\/ies that the Schr\"odinger operator $L_S$ is
invariant under the operators~(4) since
\setcounter{equation}{4}
\renewcommand{\theequation}{\arabic{equation}}
\be
[L_S, J_{jk}] =0 =[L_S, J_{0j}], \qquad j , k= 1, 2, \ldots, n.  
\ee
The operators (4) satisfy the following commutation relations:
\setcounter{equation}{5}
\renewcommand{\theequation}{\arabic{equation}{\rm a}}
\be
[J_{jk}, J_{qr}]=i\hbar (\delta_{rk} J_{jq} -\delta_{qk} J_{jr} +
\delta_{qj} J_{kr} -\delta_{jr} J_{kq}),
\ee
\setcounter{equation}{5}
\renewcommand{\theequation}{\arabic{equation}{\rm b}}
\be
[J_{jk}, J_{0q}] =i\hbar (\delta_{qj} J_{0k} -\delta_{kq} J_{0j}),
\ee
\setcounter{equation}{5}
\renewcommand{\theequation}{\arabic{equation}{\rm c}}
\be
[J_{0j}, J_{0k}] =-i\hbar \frac{f f'}{p} J_{jk},
\ee
$(j,k,q,r = 1,2,\ldots, n \geq 2)$ where $\ds f' \equiv f'(p)
=\frac{d f(p)}{d p}$. For $f(p) = p$,
relations (6) are commutation relations of the Lotentz Lie algebra [4].
When
$f(p)\not= p$, the invariance relations (5) are still valid. In this
case, the operators $\{J_{jk}, J_{0q}; j,k,q =1,2,\ldots,n\}$ still
form a Lie algebra isomorphic to the Lorentz algebra when $f
=\sqrt{p^2+\mbox{const}}$. For other choices of $f$, the operators
$\{J_{jk}, J_{0q}\}$ do not form a Lie algebra although it is
possible that they could be embedded in a Lie algebra of larger
dimension than the Lorentz algebra. We will not consider this
approach in the present paper. 

Using the symmetries (5), one can derive fundamental solutions $\psi$ of
$L_S \psi = 0$ which are invariant under the operators (4) by
requiring that $J_{jk} \psi = 0$,
$J_{0q}\psi = 0$ $(j, k, q = 1,2, \ldots, n)$. In terms of the
Fourier transform $\widetilde \psi $ of $\psi$ , this procedure leads
to: 
\setcounter{equation}{6}
\renewcommand{\theequation}{\arabic{equation}}
\be
\widetilde{\psi} (k,t) =c f(k)^{-1/2} \exp \left(-\frac{itk^2}{2m
\hbar}\right),
\ee 
where $\ds k \equiv \left(\sum_{j=1}^n k_j k_j\right)^{1/2}$ and $c$
denotes a complex constant. This result was f\/irst obtained (for the
case $f(k)=k$) in [13]. In order to calculate the inverse Fourier 
transform of (7), it is convenient to specify the function $f$. A
convenient set of choices is:

\medskip

\noindent
{\bfseries \itshape Case I:} $f(k) = k^\alpha$, \ $0 < \alpha < 2n$.

\smallskip

This reduces to the Lorentz case when $\alpha = 1$. One f\/inds from
(7) (see [18] for a derivation in the case $\alpha = 1$):
\be
\ba{l}
\ds \psi (x,r) = (2\pi \hbar)^{-\frac n2} \int_{{\mathbb R}^n} \exp
\left(\frac{i}{\hbar} k\cdot x\right) \widetilde \psi(k,t) \, d^n k
\vspace{3mm}\\
\ds \qquad = \left( \frac{m}{2\pi i \hbar t}\right)^{\frac{n}{2}
-\frac{\alpha}{ 4}} \frac{\Gamma \left( \frac n2 -\frac n4
\right)}{\Gamma\left( \frac n2 \right)} \; {}_1F_1 \left( \frac n2 -\frac
\alpha 4; \frac n2;\frac{im|x|^2}{2\hbar t}\right)
\ea
\ee
(valid in all dimensions $n\geq 2$ if $0 < \alpha < 2n$) in terms of
the conf\/luent hypergeometric function ${}_1F_1$, and we have chosen
the constant $c$ in (7) so that (8) reduces to the standard
Galilei-invariant fundamental solution 
\be
\psi_g (x,t) =\left( \frac{m}{2\pi i\hbar t}\right)^{\frac n2}
\exp\left(\frac{im|x|^2}{2\hbar t}\right)
\ee
when $\alpha = 0$. For the case $n = 3$ and $\alpha = 1$, one may use
the recurrence relations for the conf\/luent functions and the
well-known relations 
\[
{}_1F_1\left( \nu+\frac 12; 2\nu +1;
2iz\right)=\Gamma(1+\nu)e^{iz}\left(\frac z2\right)^{-\nu} J_\nu(z)
\]
to write (8) in terms of Bessel functions (see [13, 3]):
\be
\psi(x,t) =\frac{i^{\frac 14} \pi^{\frac 34}}{2}
\left(\frac{m}{2\pi i \hbar t}\right)^{\frac 32} \exp
\left( \frac{im|x|^2}{4\hbar t}\right) |x|^{\frac 12}
\left[ J_{- \frac 14}  \left( \frac{m|x|^2}{4\hbar t}\right)
+iJ_{\frac 34} \left( \frac{m|x|^2}{4\hbar t}\right)\right],
\ee
Similarly, when $\alpha =n\geq 2$ the same relation between ${}_1F_1$, and
$J_\nu$ given above can be used to write (8) in the following Bessel
function form: 
\[
\psi(x,t) =\frac{\sqrt{\pi}} {2^{\frac n2 -1}}
\left(\frac{m}{2\pi i \hbar t}\right)^{\frac n4} \exp
\left( \frac{im|x|^2}{4\hbar t}\right) \left( \frac{m|x|^2}{8\hbar
t}\right)^{- \left(\frac n4-\frac 12\right)} 
J_{\frac n4 -\frac 12} \left( \frac{m|x|^2}{4\hbar t}\right),
\]
where the following well-known relation between $\Gamma$ functions
has also been used:
\[
\Gamma(z) \Gamma\left(z+\frac 12\right) =
\frac{\sqrt{\pi} \Gamma(2z)}{2^{2z-1}}.
\]
We note that, when $n$ is a
multiple of 4, the Bessel function in the above expression can be
further simplif\/ied to a combination of sine, cosine, and algebraic
functions. 

The fact that the fundamental solutions (8) have a slower time decay,
$t^{-\left(-\frac n2 -\frac\alpha 4\right)}$ than the corresponding
Galilean result (9), which is seen to be $t^{-\frac n2}$, was 
discussed in [18] for the case $\alpha=1$. The point was made that,
although this property makes (8) unacceptable for use in quantum
mechanics (when used as a kernel analogously to the usual use of the
Galilean f\/undamental solution (9)) because the standard probability
interpretation for the wave functions of free nonrelativistic
particles is not obtained; its use may actually be more desirable
than (9) for mathematical applications because of the smoothing
properties which the linear operators that are constructed from them
possess. (Analogous operators constructed from (9) do not have such
smoothing properties.) 

One advantage of considering the cases $\alpha\not=1$
of Case I is that the corresponding convolution mappings formed
from (8) or (10) have improved smoothing properties as $\alpha$ increases.
Thus, generalizing the discussion in [18] for the case $\alpha = 1$,
we def\/ine the mappings $G_n:  g \to  G_n(\alpha)g$:
\[
(G_n(\alpha)g)(x,t) = (2\pi \hbar)^{-\frac n2}
\int_{{\mathbb R}^n}  \exp\left( \frac i\hbar k \cdot  x\right)
\left(|k|^2\right)^{-\frac \alpha 4} \exp\left(-\frac{it |k|^2}{2m\hbar}
\right) \widetilde g (k)\,  dk,
\]
and deduce that the maps $G_n(\alpha)$
are smoothing in the sense that $G_n g$ have $\ds \frac \alpha 2$
(distribution) derivatives if $g \in L^2 \left({\mathbb R}^n\right)$
$(n\geq 2)$.

The above argument shows that the smoothing properties of mappings on
$L^2$  constructed with the fundamental solutions (8) or (10)
increase as $\alpha$ increases. However, since
$f(p)= p^\alpha$ of Case I only increases algebraically with $\alpha$,
and $\alpha$ is bounded above by~$2n$, the smoothing properties are limited.
This suggests that one consider functions $f$ of exponential type in
the variable $k^2$:

\medskip

\noindent
{\bfseries \itshape Case II:}      $f(k) =\exp \left(2\beta |k|^2\right)$,
\ $(n\geq 2)$.

\smallskip

In this case, the corresponding mappings $G_n$ have much improved
smoothing properties because $G_n: \ L^2\left({\mathbb R}^n\right)
\to L^q\left({\mathbb R}^n\right)$ for all $q \in  [2, \infty)$.
In addition, it turns out that the corresponding fundamental solutions
(i.e., the inverse Fourier transforms of (7)) have the same
asymptotic time decay as $t \to +\infty$ as the Galilean fundamental
solutions (9):
\be
\psi (x,t)= c\left( 2\hbar \beta +\frac{it}{m}\right)^{-\frac n2} \exp
\left( -\frac{|x|^2}{4\hbar^2\left( \beta+\frac{it}{2m\hbar}\right)}\right).
\ee
The fact that the functions (11) have the same asymptotic decay as
the Galilean fundamental solutions (9) is explained by the fact
that the former correspond to imaginary time translations of the latter:
$\psi(x,t) = \psi_g(x,t - 2m i\hbar \beta)$ with the choices
$c= (2\pi \hbar)^{-\frac n2}$ for the constant in (11).
The increase in smoothness of the mappings $G_n$
associated with (11) relative to those associated with (9)
is intimately connected with the fact that the time translations
leading from (9) to (11) are imaginary.

\subsection*{Free-particle wave equations for arbitrary spin}

According to the principles of quantum mechanics, a wave function which
represents a free nonrelativistic particle with spin
$\ds s = \frac 12, 1, \frac 32, \ldots$ should have $2s +1$
components. The equations obtained by Hurley [5]:
\be
L_H \psi \equiv
\left(
\begin{array}{ccc}
i\hbar \p_t I_{2s+1} & \ds -\frac 1{\hbar s}S \cdot p &
\ds -\frac 1{\hbar s}K^* \cdot p
\vspace{3mm}\\
\ds \frac{1}{2m \hbar s}S \cdot p & I_{2s+1} & 0_{2s+1, 2s-1}
\vspace{3mm}\\
\ds \frac{i}{2m \hbar s}K \cdot p & 0_{2s-1, 2s+1} & I_{2s-1}
\end{array}
\right)
\left( \begin{array}{c} \psi \\ \chi \\ \Omega \end{array}\right)=0,
\ee
are a system of $6s+1$ equations in which the f\/irst row in (12)
gives the equations of motion for the $(2s +1)$-component
wave function $\psi$ and the remaining two rows are equations of
constraint def\/ining the redundant functions $\chi$ ($2s +1$
components) and $\Omega$ ($2s-1$ components) in terms of
$\psi$. $S_j$ and $K_j$ $(j = 1,2,3)$
are spin matrices with dimensions $(2s+1)\times (2s-1)$,
$(2s -1)\times (2s +1)$, respectively, and can be chosen to
satisfy the following relations [5] (with $K^*$ the adjoint of
$K$ and $\varepsilon_{ijk}$
the completely antisymmetric Levi-Civita symbol):
\be
S_i S_j +K^*_i K_j =is\hbar \varepsilon_{ijk} S_k +\hbar^2
s^2 \delta_{ij}.
\ee
The symbol $I_m$ denotes the $m$-dimensional unit matrix, and $0_{ab}$
denotes the zero matrix with $a$ rows and $b$ columns. Equations
(12) reduce to the equations derived by Levy-Leblond [19] when
$\ds s = \frac 12$ and by Hagen [20] when $s = 1$. They are Galilei
invariant by construction. For convenience in the demonstration of
nonlocal invariance of equations~(12) in the discussion to follow,
we have used a dif\/ferent normalization than Hurley.

By substituting the equations of constraint (the second and third rows
of (12)) into the equations of motion for $\psi$
(f\/irst row of (12)), and using (13), one f\/inds that each component of
$\psi$ satisf\/ies equation~(1). In order to establish the nonlocal
symmetry of equations~(12), we def\/ine new representations of
the commutation relations (6). Thus, in place of the operators (4),
we def\/ine (with $j,k= 1,2,3$):
\setcounter{equation}{13}
\renewcommand{\theequation}{\arabic{equation}{\rm a}}
\be
\widetilde{J}_{jk}=(x_j p_k -x_k p_j) I_{6s+1}-\frac 1m
(\lambda_j p_k -\lambda_k p_j),
\ee
and
\setcounter{equation}{13}
\renewcommand{\theequation}{\arabic{equation}{\rm b}}
\be
\widetilde{J}_{0j} = \frac{1}{2m}(f(p)\widetilde G_j +
\widetilde G_j f(p)),
\ee
where
\setcounter{equation}{13}
\renewcommand{\theequation}{\arabic{equation}{\rm c}}
\be
\widetilde G_j  = (tp_j -mx_j)I_{6s+1}+\lambda_j,
\ee
with
\setcounter{equation}{13}
\renewcommand{\theequation}{\arabic{equation}{\rm d}}
\be
\lambda_j=
\left(
\begin{array}{ccc}
0_{2s+1,2s+1} & 0_{2s+1,2s+1} & 0_{2s+1,2s-1}\\[2mm]
\ds \frac{1}{2s}S_j & 0_{2s+1,2s+1} & 0_{2s+1,2s-1}\\[2mm]
\ds \frac{1}{2s}K_j & 0_{2s-1,2s+1} & 0_{2s-1,2s-1}
\end{array}
\right).
\ee
Then, using (13), one verif\/ies that solutions of (12) are invariant
under the operators (14) in the sense that $[L_H, \vartheta]\psi = 0$
when $L_H \psi = 0$, with $\vartheta =\widetilde J_{jk}, \widetilde J_{0j}$
$(j,k= 1,2,3)$ and that $\{ \widetilde J_{jk}, \widetilde J_{0j};
j, k = 1, 2, 3\}$ satisfy the commutation relations (6)
with $\{ J_{jk}, J_{0j}\}$ replaced by $\{\widetilde J_{jk},
\widetilde J_{0j}\}$, respectively.

We can now construct solutions of (12) in the form
\setcounter{equation}{14}
\renewcommand{\theequation}{\arabic{equation}}
\be
\psi =  \; \mbox{column}\; \left( \psi , -\frac{i}{2m \hbar s}S\cdot p\psi,
-\frac{i}{2m\hbar s}K \cdot p \psi\right)
\ee
with $\psi(x,t) = \beta(t) F(x,t)$,
where $\beta (t)$ has $2s +1$ components and $F(x,t)$ is a (scalar)
solution of (1). Then, inserting (15) into (12) and using (13),
we infer that the components of $\beta$
 must be constant. In particular, $F(x,t)$ can be taken as the
 invariant solution (10). However, in order to write the solutions (15)
 in complete detail, one needs representations for the spin matrices
 $S_j$ and $K_j$ $(j = 1,2,3)$. Examples of these are given in [5].

Finally, we note that the operators $\widetilde J_{0j}$
def\/ined in (14b)--(14d) are pseudodif\/ferential operators and
that the terms $\ds -\frac 1m (\lambda_j p_k -\lambda_k p_j)$
$(j,k=1,2,3)$ in (14a), which play the role of spin angular momenta
in the present case, are noncanonical in the sense that they do not
satisfy the usual angular momentum commutation relations.
Thus, they do not generate the usual representations of the
three-dimensional rotation group corresponding to spin $s$.
Nevertheless, as we noted above, the operators
$\{\widetilde J_{jk}, \widetilde J_{0j}; j,k = 1,2,3\}$
satisfy the correct commutation relations.

\section{Nonfree equations possessing exact
nonlocal Lorentz symmetries}

In this section we discuss how the exact nonlocal symmetry of
free-particle equations, discussed in the preceding section,
can be extended to some nonfree equations which describe interactions
between particles. The existence of nonlocal symmetries for these
equations follows from the fact that appropriate Lie algebras
can be constructed which are isomorphic to the Lie algebra formed
by the operators in equations (6) with $\ds \frac{ff'}{p} = 1$.
We discuss this situation for two cases: (1) linear
Schr\"odinger equations with linear and quadratic potentials
(with arbitrary time-dependent coef\/f\/icients), and (2) some
classes of nonlinear Schr\"odinger equations.

\subsection*{Linear
Schr\"odinger equations with linear and quadratic potentials}

The symmetry properties of these equations have been well-studied,
especially in one space dimension. We refer to a recent
discussion of the latter case in the context of coherent states and
squeezed states [21], from which many references may be traced.

Our approach to these equations is based on the fact that they can be
transformed to the free Schr\"odinger equation (1).
The results are analogous to those obtained by Niederer~[9] for
linear Schr\"odinger equations with time-independent harmonic
oscillator potentials which showed that the Lie algebras of
the oscillator and Schr\"odinger groups are isomorphic.
Our transformation results generalize the treatment of Niederer
and are a direct extension of those of Truax [22], Bluman [23],
and of Bluman and Shtelen~[24] for the case of one spatial dimension.
However, our objective is dif\/ferent than that of the authors cited
above in that we are interested in algebras of nonlocal
symmetries rather than in algebras of point transformations.

For Schr\"odinger equations of the form:
\be
i\hbar \p_t \psi(x,t) +\frac{\hbar^2}{2m} \Delta \psi (x,t)=
\left( a(t)|x|^2 +b_j(t) x_j +c(t)\right)\psi (x,t),
\ee
with $\ds |x|^2=\sum\limits_{j=1}^n x_j x_j$ for $n \geq 1$,
where $a$, $b_j$ $(j = 1, 2,\ldots, n)$, and $c$ are arbitrary
functions of $t$, we def\/ine the following transformation:
\setcounter{equation}{16}
\renewcommand{\theequation}{\arabic{equation}{\rm a}}
\be
\psi(x,t)=\exp \left( -\frac i\hbar \left( A(t)|x|^2+B_j(t)x_j+
C(t)\right)\right)u(y,\tau),
\ee
with
\setcounter{equation}{16}
\renewcommand{\theequation}{\arabic{equation}{\rm b}}
\be
\tau(t)=\int^t \sigma^2(\mu) \, d\mu, \qquad
y_j =\sigma(t)x_j +\rho_j(t),
\ee
where $\sigma(t)$, $A(t)$, $C(t)$, $\rho_j(t)$, and $B_j(t)$
$(j = 1,2,\ldots, n)$ are to be expressed in terms
of $a(t)$, $b_j(t)$ $(j = 1, 2,\ldots, n)$, and $c(t)$.
By substitution of (17) into (16), one can choose the former
coef\/f\/icients so that (16) reduces to equation (1) for the function
$u(y,\tau)$:  
\[
 i\hbar \p_\tau u(y,\tau)+\frac{\hbar^2}{2m}\Delta _y
u(y,\tau) = 0.
\]

For the standard harmonic oscillator:
$\ds a = \frac 12 m\omega^2$, $b_j = 0$ $(j = 1, 2,\ldots,n)$,
$c = 0$; (17) reduces to the transformation obtained by Niederer:
\setcounter{equation}{17}
\renewcommand{\theequation}{\arabic{equation}}
\be
\psi (x,t)=\left(\sec (\omega t)\right)^{\frac n2} \exp
\left(-\frac{im \omega}{2\hbar}\tan (\omega t)|x|^2\right)u(y,\tau),
\ee
with $y_j=\sec (\omega t)x_j$ $(j = 1, 2,\ldots, n)$
and $\ds \tau =\frac 1\omega \tan (\omega t)$. By taking $u(y, \tau)$
to be the Galilean fundamental solution (9) expressed in the
$y$, $\tau$ variables, one obtains the standard fundamental
solution for the harmonic oscillator ([9] and [25], p.63).
In a similar manner, we can use the mapping (18) to obtain 
nonlocal invariant solutions by taking for $u(y,\tau)$ the
nonlocal invariant solution (8) of equation (1):
\[
u(y,\tau)=\left( \frac{m}{2\pi i\hbar \tau}\right)^{\frac n2 -\frac \alpha 4}
\frac{\Gamma \left(\frac n2 -\frac \alpha 4\right)}{\Gamma \left( \frac n2
\right)} \;{}_1F_1 \left(\frac n2 -\frac \alpha 4; \frac n2; \frac{im|y|^2}
{2\hbar \tau}\right), \qquad n\geq 2,
\]
and obtain
\[
\ba{l}
\ds \psi (x,t)=\left( \frac{m\omega}{2\pi i\hbar \sin(\omega t)}\right)^{\frac n2}
\left( \frac{m\omega \tan (\omega t)}{2\pi i\hbar}\right)^{-\frac \alpha 4}
\frac{\Gamma \left( \frac n2 -\frac \alpha 4\right)}{\Gamma \left(\frac n2
\right)}
\vspace{3mm}\\
\ds \qquad \times \exp \left( -\frac{im\omega}{2\hbar}\tan (\omega t)|x|^2
\right)\;{}_1F_1\left( \frac n2 -\frac \alpha 4; \frac n2;
\frac{im \omega |x|^2}{\hbar \sin(2\omega t)}\right).
\ea
\]

\noindent
{\bf Remark.}
 An explicit form for the generators of the Lorentz algebra for the
harmonic oscillator is as follows:
\[
\hat J_{0j}=\frac 1{2m\omega} \left( \hat p \hat G_j +\hat G_j \hat p\right),
\qquad \hat p\equiv \left( \sum_{j=1}^n \hat p_i \hat p_i\right)^{\frac 12},
\quad n\geq 2,
\]
where
\[
\hat p_j =\cos (\omega t) p_j -m\omega \sin (\omega t)x_j,
\]
and
\[
\hat G_j =-\sin (\omega t)p_j -m\omega \cos (\omega t) x_j,
\qquad j=1,2,\ldots, n,
\]
in terms of the same coodinates $x_j$, and momentum operators
$p_j$ used in Section~2.
We use the same angular momentum operators $J_{ik}$
$(j, k = 1,2,\ldots,n)$  as def\/ined in (4a) and obtain the following
commutation relations in place of (6b), (6c) for the free
case:
\[
\ba{l}
[J_{jk}, \hat J_{0q}]= i\hbar (\delta_{qj}\hat J_{0k}-\delta_{qk}\hat
J_{0j}),\\[2mm]
[\hat J_{0j}, \hat J_{0k}]= -i\hbar J_{jk}, \qquad
  j, k = 1, 2, \ldots, n\geq  2.
\ea
\]

We derive similar results for linear potentials by setting $a = 0$,
$c =0$, $ b_j\not= 0$ $(j = 1,2,\ldots ,n\geq 1)$ and obtain mappings
of solutions of (1) onto solutions of (16): 
\[
\ba{l}
\ds \psi(x,t)=\exp \left( -\frac i\hbar \left(tb_j x_j
+\frac{|b|^2t^3}{6m}\right) \right) u(y,\tau); 
\vspace{3mm}\\
\ds y_j=\sigma x_j +\frac{\sigma t^2}{2m} b_j, \qquad
(j=1,2,\ldots, n) \qquad \tau =\sigma^2 t.
\ea
\]
$\sigma= \mbox{const}$; and mappings of nonlocal invariant
fundamental solutions (with $n \geq  2$): 
\[
u(y,\tau)=\left( \frac{m}{2\pi i\hbar \sigma^2 t}\right)^{\frac n2 -
\frac \alpha 4} \frac{\Gamma \left(\frac n2 -\frac \alpha 4\right)}
{\Gamma \left( \frac n2 \right)} \;{}_1F_1 
\left(\frac n2 -\frac \alpha 4; \frac n2; \frac{im|x|^2}
{2\hbar t} +\frac{it}{2\hbar} b_j x_j + \frac{i|b|^2 t^3}{8m \hbar}
\right), 
\]
\[
\ba{l}
\ds \psi(x,t)=\left( \frac{m}{2\pi i\hbar \sigma^2 t}\right)^{\frac n2 -
\frac \alpha 4} \frac{\Gamma \left(\frac n2 -\frac \alpha 4\right)}
{\Gamma \left( \frac n2 \right)} 
\exp\left( -\frac i\hbar \left(tb_j x_j +\frac{|b|^2 t^3}{6m}\right)
\right)
\vspace{3mm}\\
\ds \qquad \times \; {}_1F_1 
\left(\frac n2 -\frac \alpha 4; \frac n2; \frac{im|x|^2}
{2\hbar t} +\frac{it}{2\hbar} b_j x_j + \frac{i|b|^2 t^3}{8m \hbar}
\right), 
\ea
\]
Similar results can be given for the case of a harmonic oscillator
driven by an external force $f_j(t)$ $(j = 1,2,\ldots, n)$, 
for which $\ds a = \frac 12 m\omega^2 = \mbox{const}$ and $b_j=-f_j$,
whose Galilean fundamental solutions were given in [25]. However,
since the expressions obtained are somewhat unwieldy, we shall not
give those results here. 

\subsection*{Nonlinear Schr\"odinger equations with exact nonlocal
symmetry} 

Doebner and Goldin [6, 7] considered a class of nonlinear Schr\"odinger
equations which were suggested by their studies of dissipative
quantum theory based on group theoretic considerations relating to
groups of dif\/feomorphisms on Euclidean spaces and the corresponding
Lie algebras (algebras of vector f\/ields). Related equations were
considered earlier by Sabatier [26] and subsequently by Auberson and
Sabatier [8] and by Auberson [27]. 

We f\/irst consider a subclass of the Doebner-Goldin (DG) equations
which are linearizable in the sense that they can be mapped to linear
Schr\"odinger equations by point transformations. The nonlocal symmetry
of the linearizable DG equations follows from the isomorphy of the
Lie algebras of these equations with the nonlocal Lie algebras of
linear Schr\"odinger equations. 

The DG equations can be written in the form:
\be
i\hbar \psi_t =\left( -\frac{\hbar^2}{2m} \Delta +V(x,t)\right) \psi
+\frac{i\hbar D}{2} R_2(\psi, \overline{\psi})\psi +\hbar D'\sum_{j=1}^5
c_j R_j(\psi, \overline{\psi}) \psi,
\ee
where $D$ and $D'$ denote constant dif\/fusion coef\/f\/icients, and the
real-valued nonlinear functionals $R_j(\psi, \overline \psi)$ $(j =
1,2,3,4,5)$ are given by: 
\be
\ba{l}
\ds R_1(\psi, \overline \psi)=\frac{\nabla \cdot \widetilde j}{\rho},
\qquad
R_2(\psi, \overline \psi)=\frac{\Delta \rho}{\rho},
\qquad 
R_3(\psi, \overline \psi)=\frac{\widetilde j^2}{\rho^2},
\vspace{3mm}\\
\ds 
R_4(\psi, \overline \psi)=\frac{\widetilde j \cdot \nabla
\rho}{\rho^2}, 
\qquad
R_5(\psi, \overline \psi)=\frac{(\nabla \rho)^2}{\rho^2},
\ea
\ee
where $\rho =\overline \psi \psi$, and $\widetilde j$ is related to
the usual probability current density $j$ by  
\[
\widetilde j=\frac m\hbar j =\frac 1{2i} (\overline \psi\nabla
\psi -\psi \nabla \overline \psi).
\]
Lie symmetry analyses of these
equations have been discussed by several authors [28--30]. References
to discussions of analogous equations by other authors can be found
in [7]. 

It was pointed out by Doebner and Goldin that the subfamily of their
equations (19), (20) def\/ined by the following relations between the
coef\/f\/icients $D$, $D'$, $c_j$ $(j=1,2,3,4,5)$:
\be
D=D'c_1=-D'c_4; \qquad D'(c_2+2c_5)=D'c_3=0,
\ee 
are linearizable and that the corresponding solutions can be
constructed from solutions of linear Schr\"odinger equations by means
of ``nonlinear gauge transformations'' (see also~[28]).  The
nonlinear gauge transformations of DG are given by: 
\be
\psi \to \psi' =N(\psi) =|\psi|\exp\biggl(i[\gamma \ln |\psi|+\Lambda
\,\mbox{Arg}\, \psi]\biggr),
\ee
with $\gamma$, $\Lambda$ real numbers (and $\Lambda \not =0$).
Doebner and Goldin show that, given the relations~(21) among the
coef\/f\/icients, if $\psi$ is a solution of (19), (20); then $\psi'=N(\psi)$
is a solution of the following linear Schr\"odinger equation: 
\be
\frac{i\hbar}{\Lambda} \psi_t' =-\frac{\hbar^2}{2m \Lambda^2}
\Delta \psi ' +V(x,t) \psi',
\ee
when
\[
\Lambda=\left( 1-\frac{4m}{\hbar} D' c_2 -\frac{4m^2 D^2}{\hbar^2}
\right)^{-\frac 12}
\]
and
\be
\gamma=-\frac{2m D\Lambda}{\hbar},
\ee
provided that $\ds \frac{4m}{\hbar} D' c_2 + \frac{4m^2 D^2}{\hbar^2} <1$.
Since the gauge transformations satisfy the
group law $N_{\gamma_1, \Lambda_1} \bullet N_{\gamma_2, \Lambda_2}
= N_{\gamma_1+\Lambda_1 \gamma_2, \Lambda_1\Lambda_2}$,
the gauge transformations inverse to (22) are given by:
\be
\psi =N^{-1} (\psi') =|\psi'| \exp
\left( i\left[ -\Lambda^{-1}\gamma \ln |\psi'| +
\Lambda^{-1} \mbox{Arg}\, \psi'\right]\right).
\ee
This mapping is analogous to the mapping (17) which 
transforms solutions of the free Schr\"odinger equation to 
solutions of the linear Schr\"odinger equation with linear 
or quad\-ra\-tic potentials.

For the case when the potential $V$ is identically zero, $\psi'$
is a solution of the free Schr\"odinger equation with the mass 
$m$ replaced by the ``ef\/fective mass'' $m\Lambda$. 
For example, we may take the solution analogous to (8):
\be
\psi'(x,t)= \left( \frac{m\Lambda}{2\pi i\hbar t}\right)
^{\frac n2 -\frac \alpha 4} 
\frac{\Gamma \left( \frac n2 -\frac \alpha 4\right)}{\Gamma \left(
\frac n2\right) } \; {}_1F_1\left( \frac n2 -\frac \alpha 4;
\frac n2; \frac{im \Lambda|x|^2}{2\hbar t}\right)
\ee
for $0 < \alpha < 2n$ and $n \geq  2$. 
More generally, one can also consider mappings from solutions of 
the DG equations (19), (20) to solutions of a linear Schr\"odinger 
equation with one of the potentials discussed in the f\/irst part 
of this section and then map to solutions of the free Schr\"odinger
equation by using the results discussed there. The composition of 
this sequence of mappings yields transformations of solutions 
of (19), (20) to solutions of free Schr\"odinger equations 
without the necessity of assuming that the linear potential 
in the~DG equations is identically zero.

Auberson and Sabatier [8] (AS) considered the following NSE 
(for convenience, we set $\hbar = 1$ and $2m = 1$):
\be
i\psi_t(x,t) =(-\Delta +V) \psi(x,t) +s
\frac{\Delta |\psi|}{\psi} \psi(x,t),
\ee
where $s$ is a real parameter. For $s < 1$ AS use the following 
linearization transformation:
\be
\psi=|\psi|\exp(-i \theta), \qquad
t=(1-s)^{-\frac 12} t',
\qquad \theta (x,t)=(1-s)^{\frac 12} \theta'(x,t'),
\ee
which transforms equation (27) to the following 
linear Schr\"odinger equation:
\[
i\psi'_{t'} (x,t') =-\Delta \psi'(x,t') +(1-s)^{-1} V(x) \psi'(x,t')
\]
for the quantity
\be
\psi'(x,t') =\left| \psi\left(x,(1-s)^{-\frac 12} t'\right) \right|
\exp(-i\theta'(x,t')).
\ee
AS also linearize (27) when $s\geq 1$, 
but the linear equations thereby obtained are
not Schr\"odinger equations so we shall not discuss them.

If we consider invariant solutions analogous to (26) for (29):
\[
\psi'(x,t') =(4\pi i \hbar t')^{-\left(\frac n2-\frac \alpha 4\right)}
\frac{\Gamma\left(\frac n2 -\frac \alpha 4\right)}{\Gamma \left(
\frac n2\right)} \; {}_1F_{1} \left( \frac n2 -\frac \alpha 4; \frac n2;
\frac{i|x|^2}{4t'} \right),
\]
then the corresponding invariant solutions of (27) are obtained from (28).

As a consequence of the analogies between the inverse gauge 
transformation (25) of the DG equations or of the linearization transformation 
(29) of the AS equation with the transformation (17) between solutions 
of the free Schr\"odinger equation and solutions of linear 
Schr\"odinger equations with linear or quadratic potentials, 
we see that the symmetry algebras corresponding, respectively, to
the free Schr\"odinger and to the linearizable DG or AS equations 
are isomorphic.

\section{Asymptotic symmetry}

For general nonlinear Schr\"odinger equations, one does not 
expect the nonlocal symmetries for the equations discussed in 
the preceding sections to extend in an exact form because those 
results depended on the fact that the equations were
either linear or linearizable. Lie's algorithm [3, 11] gives a 
general method for the investigation of local symmetries and 
solutions of dif\/ferential equations, including nonlinear ones; 
and this method has been applied to many types of equations, both
linear and nonlinear. However, as we have noted in the Introduction, 
since the  symmetries that we are discussing are nonlocal and are 
def\/ined in terms of pseudodif\/ferential, rather than 
dif\/ferential, operators; Lie's approach and related techniques
are not adequate to deal with them. Because of this dif\/f\/iculty of 
extending nonlocal symmetries to general nonlinear equations, 
we propose to use a definition of symmetry based on a reducibility
property (see Definition 1.2 and the discussion below).
When there is no exact reducibility, we
introduce a weaker concept of {\it asymptotic symmetry}.

\subsection*{Power-type nonlinearities}

Consider an NSE of the form
\be
i\p_t\psi +\frac 1{2m} \Delta \psi =
F\left(\psi, \overline \psi, \p_{x_j}\psi,
\p_{x_j} \overline \psi, \p^2_{x_ix_j}\psi, \p^2_{x_ix_j}\overline
\psi\right) 
\ee
$(i,j=l,2,\ldots,n)$, where the nonlinear function $F$ depends in general on
the solution $\psi$, its complex conjugate $\overline \psi$, derivatives of
these functions through the second order and, unless a statement is made to
the contrary, we set $\hbar  = 1$ in the present section. 

Following Def\/inition 1.2, we will say that an operator $Q$ is a symmetry of
equation~(30) if and only if the corresponding ansatz, obtained as a
solution of~(3), reduces~(30) to a system of PDEs in fewer independent
variables or, as a limiting case, to a system of ordinary dif\/ferential
equations (ODEs). This approach is especially useful when one wants to
extend symmetries of a system of linear equations 
(in the present case the free Schr\"odinger equation (1)) to a 
system of nonlinear equations.

By extending an argument in [13] for Case I with $\alpha = 1$, we conclude
that the following ansatz is invariant under the algebra (4):
\be
\psi (x,t) =\phi(t) g(t,x),
\ee
where $\phi$ is an arbitrary function of $t$ and $g$ has the form (8) (or
(10) in the special case when $n=3$ and $\alpha = 1$). 
According to Def\/inition 1.2, we say that {\it
equation (30) is invariant under
the algebra} (4) if and only if the ansatz (31) reduces (30) to 
an ODE for the function~$\phi(t)$. 
In the following def\/inition, we introduce a concept of {\it asymptotic
symmetry} when an exact reduction does not exist, but a reduction does exist
in an asymptotic sense. 

\medskip

\noindent
{\bf Def\/inition 4.1.} {\it We will say that equation (30) has the asymptotic
symmetry (4) if and only if the ansatz~(31) reduces (30) to an ODE for
$\phi(t)$ in the asymptotic region $m|x|^2 \ll 2t$.}

\medskip

In this section we will f\/irst discuss NSEs with power-type nonlinearities
and then consider several cases of derivative nonlinearities. We f\/irst treat
the case in which the nonlinear term in (30) is of the form:
\be
F=\lambda(\overline \psi \psi)^k \psi,
\ee
where $k$ denotes a positive real number (not necessarily an integer) and
$\lambda$ denotes a complex (coupling) constant. Lie symmetry analyses of
equations of the form (30), (32) have been discussed by many authors (see
[3] for a summary). In addition, many authors have investigated the
existence of solutions to such equations in various Banach and Hilbert
spaces. For dimensions $n \geq 3$, many of these results require that 
$\ds 0 < k < \frac {2n}{n-2}$ because the proofs use the Sobolev embedding
theorem. See [15] for a summary. Our results are not subject to this
restriction. 

To investigate the asymptotic symmetry of (30), (32), we look for a solution
of the form~(31) where $g(x,t)$ is a solution of the free Schr\"odinger
equation def\/ined by (8):
\be
g(x,t) =\left( \frac{m}{2\pi it}\right)^{\frac n2 -\frac \alpha 4}
\frac{\Gamma\left( \frac n2 -\frac \alpha 4\right)}{\Gamma \left( \frac
n2\right)} \; {}_1F_1\left(\frac n2-\frac \alpha 4; \frac n2;
\frac{im|x|^2}{2t} \right),
\ee
with $0 < \alpha < 2n$ for spatial dimensions $n \geq  2$. 
(If $n = 3$ and $\alpha = 1$ we may, of  course, use (10) instead.) 
Then, substituting these expressions into (30) and (32),
assuming that $\phi$ depends only on $t$,  
and using the fact that $g$ satisf\/ies the free
Schr\"odinger equation 
$\ds i\p_t g=-\frac{1}{2m} \Delta g$, 
we obtain the following equation for $\phi$:
\be
\phi_t =-i\lambda (\overline \phi \phi)^k (\overline g g)^k \phi,
\qquad k >0.
\ee
Since $\phi$ is assumed to depend only on $t$, the above derivation is only
consistent if the quantities $(\overline g g)^k$ are independent of the
spatial coordinates $x_j$ $(j = 1,2,\ldots, n)$. 
This is not true in general, so we consider the limit of large $t$ or, more
precisely,  values of the variables $x_j$ $(j = 1,2,\dots, n)$
and $t$ such that $m|x|^2 \ll 2t$. Then, using the small-argument expansions
\be
{}_1F_1(a;c;z)=1+\frac ac z +\frac{a(a+1)}{c(c+1)} \frac{z^2}{2} +\cdots
\ee
for the conf\/luent functions, we obtain the following asymptotic result for 
$\overline g g$:
\be
\overline g g \cong \left( \frac{m} {2\pi t} \right)^{n-\frac \alpha 2}
\left( \frac{\Gamma \left( \frac n2 -\frac \alpha
4\right)}{\Gamma\left(\frac n2\right)} \right)^2 \left( 1+O
\left( \left( \frac{m|x|^2}{2t}\right)^2\right) \right),
\ee
and (34) becomes, to leading order in the quantity $\ds 
\frac{m|x|^2}{2t}$, 
\be
\phi_t   \cong -i\lambda \left( \frac{m} {2\pi t} \right)^{\left(
n-\frac \alpha 2\right)k}
\left( \frac{\Gamma \left( \frac n2 -\frac \alpha
4\right)}{\Gamma\left(\frac n2\right)} \right)^{2k}
(\overline \phi \phi)^k \phi.
\ee
From the form of equation (37), we f\/ind that its solution must have the form
$\phi(t) =\beta \exp(-i\hbar (t))$ with $\beta$
 a real constant and $h(t)$ a real-valued function of $t$. 
Substitution of this expression into (37) yields an equation for 
$h(t)$ which has the following solution:   ($\overline \omega$
 a real constant)
\be
h(t)  = \lambda \left( \frac{m} {2\pi} \right)^{\left(
n-\frac \alpha 2\right)k}
\left( \frac{\Gamma \left( \frac n2 -\frac \alpha
4\right)\beta}{\Gamma\left(\frac n2\right)} \right)^{2k}
\chi_k(t) +\overline \omega,
\ee
where $\chi_k(t) =t^{-\left( n-\frac \alpha 2\right)k+1}$
 when $\ds k \not= \left(n-\frac \alpha 2\right)^{-1}$ and 
$= \ln t$ when $\ds k = \left(n-\frac \alpha 2\right)^{-1}$. The
asymptotic solution to (30), (32) is now obtained by substitution into (31):
\[
\psi(x,t)\cong \beta \exp(-i\hbar (t)) g(x,t)
\]
with $g$ given by (33) 
and $h$ given by (38).

We note that the asymptotic result for $|\psi| =(\overline \psi \psi)^{\frac
12}$ corresponds (apart from the constant $\beta$) 
to the asymptotic result for the solution $g(x,t)$ of the free
Schr\"odinger equation, whereas the asymptotic value of 
$\ds \mbox{Arg}\, \psi =\frac 1{2i} \ln\left(
\frac{\psi}{\overline{\psi}}\right)$ 
 contains  ef\/fects of the nonlinear terms (32).

Similar results can also be obtained when the nonlinear term in (30) is a
linear combination of power-type nonlinearities such as, for example:
\be
F=-a_0 \psi -a_1 (\overline \psi \psi) \psi -a_2
(\overline \psi \psi)^2 \psi,
\ee
where $a_j$, $(j=0,1,2)$ are real constants, and $a_2\not=0$. 
Lie symmetries of equations (30), (39) were discussed by Gagnon and
Winternitz [31].

\subsection*{The Doebner-Goldin and related equations}

We next consider asymptotic symmetry results for the DG equation (19),
(20) and then discuss the relationship between these results and some others
for these equations.

To show that the DG equations have asymptotic symmetry in the sense of
Def\/inition~4.1, we follow the procedure used for NSEs with power-type
nonlinities and look for a solution $\psi$ of (19), (20) of the form (31)
with $g(x,t)$ def\/ined by (33) or (10) with the appropriate powers of $\hbar$
again inserted. We obtain the following linear equation for $\psi$ by virtue
of the homogeneity property of the nonlinear functionals $R_j$
$(j = 1, 2, 3, 4, 5)$:
\be
i\hbar \phi_t =V \varphi +\frac{i \hbar D} {2} R_2(g,\overline g)\phi
+\hbar D' \sum_{j=1}^5 c_j R_j (g, \overline g) \phi.
\ee
Using the derivative relations $\ds \frac{d}{dz} \; {}_1F_1(a;c;z)=
\frac{a}{c} \;{}_1F_1(a+1;c+1;z)$ and the small-argument expansions (35) for
the conf\/luent functions, we obtain the following asymptotic results for the
$R_j(g,\overline g)$ $(j= 1, 2, 3, 4, 5)$:
\[
R_1(g,\overline g) =\left( 1-\frac {\alpha}{2n}\right)
\frac {mn}{\hbar t}\left( 1+O \left(\left(\frac{m|x|^2}{2\hbar t}\right)^2
\right)\right),
\]
\[
R_2(g,\overline g) =-\left( 1-\frac {\alpha}{2n}\right)
\frac{2\alpha}{n} \frac {m}{\hbar t}
\frac{m|x|^2}{2\hbar t}
\left( 1+O \left(\frac{m|x|^2}{2\hbar t}\right)\right),
\]
\[
R_3(g,\overline g) =\left( 1-\frac {\alpha}{2n}\right)^2
 \frac {2m}{\hbar t}
\left(\frac{m|x|^2}{2\hbar t}\right)
+O \left(\left(\frac{m|x|^2}{2\hbar t}\right)^2 \right),
\]
\[
R_4(g,\overline g) =-\left( 1-\frac {\alpha}{2n}\right)^2
\frac{2\alpha}{n(n+2)} \frac {2m}{\hbar t}
\left(\frac{m|x|^2}{2\hbar t}\right)^2
\left( 1+O \left(\frac{m|x|^2}{2\hbar t} \right)\right),
\]
\[
R_5(g,\overline g) =
\frac{\left(1-\frac \alpha{2n}\right)^2\alpha^2}
{n^2(n+2)^2} \frac {2m}{\hbar t}
\left(\frac{m|x|^2}{2\hbar t}\right)^3
\left( 1+O \left(\frac{m|x|^2}{2\hbar t}\right)\right),
\]
Thus, we see that the DG equations are asymptotically invariant
in the sense of De\-f\/i\-ni\-tion~4.1 if we consider the case of an
identically zero potential $V$ and omit  the $R_j$
functionals with $j = 2, 3, 4, 5$.
Then, solution of (40) gives the following asymptotic
result $(m|x|^2\ll 2\hbar t)$:
\be
\psi(x,t)\cong \kappa \exp\left( -iD'c_1\left(1-\frac{\alpha}{2n}\right)
\frac{mn}{\hbar}\ln (t)\right)g(x,t)
\ee
with $\kappa$ a complex constant.

For reasons of consistency, we must show that a solution $\psi$
of (19), (20)
obtained as in (25) from a solution $\psi'$
 of (23) is consistent with our asymptotic
symmetry result (41). This can be done by noting that, for the case
when the potential $V$ is identically zero, $\psi'$
 is a solution of the free Schr\"odinger equation
with the mass $m$ replaced by the ``ef\/fective mass''
$m\Lambda$ as in (26). Using the small-argument expansion (35)
and related expansions for the arctangent and logarithmic
functions that occur in $\mbox{Arg}\, \psi'$
and $\ln(|\psi'|)$, respectively, we obtain from (25):
\be
\ba{l}
\ds \psi(x,t)\cong \left(\frac{m\Lambda}{2\pi \hbar t}\right)^{\frac n2}
\frac{\Gamma\left(\frac n2 -\frac \alpha 4\right)}{\Gamma\left( \frac n2
\right)}\exp\left(
\frac{2mi D}{\hbar}\ln \left(
\frac{\Gamma\left(\frac n2 -\frac \alpha 4\right)}{\Gamma
\left( \frac n2\right)}\left( \frac{m\Lambda}{2\pi \hbar t}\right)^{\frac
n2 -\frac \alpha 4}\right)\right)
\vspace{3mm}\\
\ds \qquad \times \exp\left( -\frac{i\pi}{\Lambda} \left(
\frac n4 -\frac \alpha 8\right) +i\left( 1-\frac{\alpha}{2n}\right)
\frac{m|x|^2}{2\hbar t}\right).
\ea
\ee
We will compare this expression with the asymptotic symmetry result (41),
which can be written in the form:
\be
\ba{l}
\ds \psi_{\mbox{\scriptsize \rm asym}}(x,t)\cong 
\kappa \left(\frac{m\Lambda}{2\pi \hbar t}\right)^{\frac n2-\frac \alpha 4}
\frac{\Gamma\left(\frac n2 -\frac \alpha 4\right)}{\Gamma\left( \frac n2
\right)}
\exp\left(
-i  \left[ D' c_1 \left(1-\frac{\alpha}{2n}\right) \frac{mn}{\hbar}
\ln(t) +\right.\right.
\vspace{3mm}\\
\ds \qquad \left.\left. + \pi\left(\frac n4 -\frac \alpha 8\right)\right]\right)
\exp\left( i\left(1-\frac{\alpha}{2n}\right)
\frac{m|x|^2}{2\hbar t}\right).
\ea 
\ee
The expressions (42), (43) will be compared in the spacetime domain 
$m|x|^2\ll 2\hbar t$ by setting
\be 
\kappa =\beta\exp (-i\delta) \qquad 
\mbox{with} \quad \beta \quad \mbox{and} \quad \delta \quad \mbox{real}.
\ee
One then obtains
\be
\beta =\Lambda^{\frac n2 -\frac \alpha 4}
\ee
and the coef\/f\/icients of $\ln t$ in the exponents of (42) and (43) 
agree because of the  equality $D =D' c_1$, 
which is part of the conditions (21) of DG required for 
linearizability. Equating the constant parts of the phases of (42) 
and (43) gives:
\be
\delta=n\left(1-\frac{\alpha}{2n}\right)
\left[ \frac \pi 4\left(\Lambda^{-1} -1\right)
-\frac{mD}{\hbar} \ln \left(\frac{m\Lambda}{2\pi \hbar}
\right) \right] -\frac{2m D}{\hbar}
\ln \left( \frac{\Gamma\left(\frac n2 -\frac \alpha 4\right)}
{\Gamma \left(\frac n2\right)}\right).
\ee
Thus, the solutions of the DG equations (19), (20) obtained by
their gauge equivalence to solutions of the linear Schr\"odinger 
equation (23) with $V=0$ is consistent with the asymptotic 
symmetry result (41) provided that the constant $\kappa$ is
chosen to satisfy~\mbox{(44)--(46).}

\subsection*{Equations of Kostin and of Bialynicki-Birula and Mycielski}

We consider NSEs of the form:
\be
i\hbar \psi_t =-\frac{\hbar^2}{2m} \Delta \psi +\xi_1 \ln (\overline \psi
\psi) \psi -\frac{\xi_2}{2i} \ln \left(\frac{\psi}{\overline \psi}\right)\psi,
\ee
where $\xi_1$ and $\xi_2$ are real constants. 
The f\/irst logarithmic term $\ds \ln \left( |\psi|^2\right)\psi$
was originally proposed by Bialynicki-Birula and Mycielski [16] 
and the second logarithmic term, which involves the 
phase of $\psi$: $\ds (2i)^{-1} \ln \left(\frac{\psi}{\overline \psi}
\right)$, was f\/irst proposed by Kostin [17] in connection with studies 
of dissipation ef\/fects in quantum mechanics. It is
appropriate to discuss equation (47) in this section because Doebner 
and Goldin have shown [7] that their equation (19), (20) extends to 
include the nonlinear terms in (47) when the parameters $\gamma$ and 
$\Lambda$ in the gauge transformation (22) are time-dependent. 
Symmetry analyses of equation (47) with $\xi_2 = 0$
have been investigated (cf. [ 32, 3])
and existence results for this equation (with $\xi_2= 0$) 
when an appropriate class of linear potentials is also 
included have been summarized by Cazenave [15].

Looking for solutions of (47) in the form (31), (33); we obtain 
the following equation for $\phi$:
\be
i\hbar \phi_t =\xi_1 \ln \left( |\phi|^2 |g|^2\right) \phi -
\frac{\xi_2}{2i} \ln \left( \frac{\phi}{\overline \phi} 
\frac{g}{\overline{g}}\right)\phi.
\ee
Then, assuming that $m|x|^2 \ll 2\hbar t$  and using the expansions (35), 
we obtain (36) and
\be
\frac{g}{\overline g} \cong (-1)^{\frac n2 -\frac \alpha 4} \left(
1+O\left(\frac{m|x|^2}{2\hbar t} \right) \right).
\ee
Equation (48) becomes, to leading order in the quantity $\ds \frac{m|x|^2}{
2\hbar t}$,
\be
\phi_t \cong -\frac i\hbar \xi_1\ln \left( \left( \frac{m}{2\pi \hbar t}
\right)^{n-\frac \alpha 2} |\phi|^2\right)\phi+
\frac{\xi_2}{2\hbar} \ln \left(\frac{\phi}{\overline \phi}\right)\phi
+\frac{i\xi_2}{2\hbar} \left(\frac n2 -\frac \alpha 4\right)\pi \phi,
\ee
where the last term on the right-hand side comes from the logarithm 
in (48) and vanishes $\ds \frac n2-\frac \alpha 4$ is an even integer. 
From the form of eq.(50), we find that its
solution must have the form $\phi(t) =\beta \exp(i\delta(t))$ with $\beta$
a real constant and $\delta(t)$ a real-valued function. 
Substitution of this expression into (50) yields the following
equation for $\delta(t)$:
\be
\delta_t =-\frac{\xi_1}{\hbar} \ln \left( \beta^2
\left( \frac{m}{2\pi \hbar t}\right)^{n-\frac \alpha 2} 
\left( \frac{\Gamma \left( \frac n2 -\alpha 4\right)}
{\Gamma\left(\frac n2\right)}\right)^2\right) +\frac{\xi_2\delta}{\hbar}
+\frac{\xi_2}{\hbar}\left( \frac n4 -\frac \alpha 8\right)\pi.
\ee
The  solution of this equation has dif\/ferent forms according 
as $\xi_2$ is zero or nonzero.

\medskip

{\bfseries  \itshape Case A.} $\xi_2=0$.
\be
\ba{l}
\ds \phi(t)=\beta \exp \left( -\frac{i\xi_1 t}{\hbar} 
\left[ \ln \left( \beta^2 \left( \frac{\Gamma\left( \frac n2 -\frac \alpha 4
\right)}{\Gamma \left(\frac n2\right)} \right)^2\right) +
\left(n-\frac \alpha 2\right) 
\left( \ln \left( \frac{m}{2\pi \hbar }\right) +1 \right)\right]\right)
\vspace{3mm}\\
\ds \qquad \times \exp\left( \frac{i\xi_1}{\hbar}
\left( n-\frac \alpha 2\right) t\ln t +i\zeta\right),
\qquad \zeta= \mbox{real const}.
\ea
\ee

{\bfseries \itshape Case B.} $\xi_2\not=0$  and $\ds \frac n2-
\frac \alpha 4 \neq$  an even integer

In this case, equation (51) can be written in the form:
\[
\ba{l}
\ds \frac{d}{dt} \left( \exp \left( -\frac{\xi_2 t}{\hbar}
\right) \delta(t)\right) =\frac{\xi_1}{\hbar}
\left( n-\frac \alpha 2\right) \ln t \exp\left(-\frac{\xi_2 t}{\hbar}
\right) +\frac{\xi_2 \pi}{\hbar} \left( \frac n4 -\frac \alpha
8\right) \exp\left(-\frac{\xi_2 t}{\hbar} \right)
\vspace{3mm}\\
\ds \qquad -\frac{\xi_1}{\hbar} \ln \left(\beta^2 \left(
\frac{m}{2\pi\hbar} \right)^{n-\frac \alpha2} 
\left( \frac{\Gamma\left(\frac n2 -\frac \alpha 4\right)}{
\Gamma\left( \frac n2\right)}\right)^2\right) \exp\left( -\frac{\xi_2
t}{\hbar}\right).
\ea\!\!
\]
Integrating this equation from $t^* > 0$ to $+\infty$, 
and using integration by parts for the $\ln t$ term, we obtain
\[
\ba{l}
\ds  \delta(t^*)=
\frac{\xi_1}{\xi_2} \ln \left(\beta^2 \left(
\frac{m}{2\pi\hbar} \right)^{n-\frac \alpha2} 
\left( \frac{\Gamma\left(\frac n2 -\frac \alpha 4\right)}{
\Gamma\left( \frac n2\right)}\right)^2\right)
-\frac{\xi_1}{\xi_2} \left(n-\frac \alpha 2\right) \ln t^*-
\left( \frac n4-\frac \alpha 8\right)\pi
\vspace{3mm}\\
\ds \qquad - \frac{\xi_1}{\xi_2} \left(n-\frac \alpha 2\right)
\exp\left( \frac{\xi_2 t^*}{\hbar}\right) E_1 \left(\frac{\xi_2
t^*}{\hbar} \right)
\ea
\]
and
\be
\ba{l}
\ds  \phi(t) =\beta \exp(i\delta(t))=
\beta\exp\left(i
\frac{\xi_1}{\xi_2} \ln \left(\beta^2 \left(
\frac{m}{2\pi\hbar} \right)^{n-\frac \alpha2} 
\left( \frac{\Gamma\left(\frac n2 -\frac \alpha 4\right)}{
\Gamma\left( \frac n2\right)}\right)^2\right)
\right)
\vspace{3mm}\\
\ds  \times \exp\left(-i\frac{\xi_1}{\xi_2} 
\left(n-\frac \alpha 2\right) \ln t-
i\frac{\xi_1}{\xi_2} \left(n-\frac \alpha 2\right)
\exp\left( \frac{\xi_2 t}{\hbar}\right) E_1 \left(\frac{\xi_2
t}{\hbar}\right)-i\left( \frac n4-\frac \alpha 8\right)\pi\right) 
\ea\!\!
\ee
in terms of the exponential integrals $\ds E_1(x)=-\,\mbox{Ei}\,(-x)
= \int_x^\infty \exp (-t) t^{-1} dt$ (where the integral is
understood as a principal value integral when $x < 0$). In Case B, 
the terms involving the quantity $\ds \left( \frac n4 -\frac \alpha
8\right) \pi$ are absent when $\ds \frac n2 -\frac \alpha 4$ is an
even integer. The f\/inal asymptotic solutions to (47) are obtained to
leading order in the quantity $\ds \frac{m|x|^2}{2\hbar t}$ by
substituting (52) and (53) into the equation $\psi(x,t) = \phi(t)g(x,t)$, 
where $g(x,t)$ is given by~(33).

\section{Concluding remarks}

We have shown that nonlocal Lorentz symmetry, previously known to be
valid for free-particle Schr\"odinger equations [4, 13], is also
valid in a modif\/ied form for the Galilei-invariant linear wave
equations of Hurley [5], which describe the time evolution of free
quantum particles with arbitrary spin. In Section~3 we 
have discussed similar results for linear Schr\"odinger equations
with linear and harmonic oscillator potentials with arbitrary
time-dependent coef\/f\/icients. For nonlinear
equations, we have shown that a subset of the nonlinear Schr\"odinger
equations introduced by Doebner and Goldin [6, 7] in connection with
studies of dissipative ef\/fects in quantum mechanics as well as some
of the related equations discussed by Auberson and Sabatier [26, 8]
have exact forms of these symmetries. Moreover, we have also shown
that several classes of nonlinear Schr\"odinger equations  -- 
those with power-type nonlinearities as well as the full set of
equations proposed by Doebner and Goldin -- have asymptotic nonlocal
symmetry in the sense described in the present paper.

The question of the nonlocal symmetry of nontrivial (i.e.,
nonfree) multiparticle Schr\"o\-din\-ger equations is open. In general, we
expect that the concept of nonlocal symmetry may be helpful in the analysis
of such equations.

\subsection*{Acknowledgments}

We wish to express our thanks, respectively, to G.~Bluman and
G.A.~Goldin
(VMS) and to
T.L.~Gill and G.A.~Goldin (WWZ) for valuable discussions concerning this
work. In addition, we thank two referees for constructive remarks concerning
an earlier version of the paper.

The initial stage of the work of the f\/irst author was partially supported by
the U.S. Army High Performance Computing Research Center under the auspices
of the Department of the Army, Army Research Of\/f\/ice cooperative agreement
number DAAH04--95--2--0003/contract number 
DAAH04--985--C--0008, the content of which does not necessarily ref\/lect the
position or the policy of the U.S. government, and no of\/f\/icial endorsement
should be inferred.

 \label{zachary-lp}

\end{document}